\documentclass{iopart}
\usepackage{iopams}  
\usepackage{graphicx}
\usepackage{todonotes}
\usepackage{mathpple} 
\usepackage{amsfonts}
\usepackage{array}
\usepackage{hyperref}
\usepackage{verbatim}
\usepackage{subcaption}

\begin{document}

\title{A detailed heterogeneous agent model for a single asset
  financial market with trading via an order book}

\author{Roberto Mota Navarro  and Hern\'an Larralde}
\address{Instituto de Ciencias F\'isicas, 
Universidad Nacional Aut\'onoma de M\'exico, Cuernavaca, 
Morelos, C.P. 62210, M\'exico}

\ead{mvr@fis.unam.mx}

\begin{abstract}
We present an agent based model of a single asset financial market
that is capable of replicating most of the non-trivial statistical
properties observed in real financial markets, generically referred to
as stylized facts. In our model agents employ strategies inspired on
those used in real markets, and a realistic trade mechanism based on
a double auction order book. We study the role of the distinct
types of trader on the return statistics: specifically, correlation
properties (or lack thereof), volatility clustering, heavy tails, and
the degree to which the distribution can be described by a
log-normal. Further, by introducing the practice of ``profit taking'',
our model is also capable of replicating the stylized fact related to
an asymmetry in the distribution of losses and gains.

\end{abstract}

\maketitle

\section{Introduction}

In the past five decades a great number of time series of prices of
various financial markets have become available and have been
subjected to analysis to characterize their statistical properties
\cite{Pagan1996,Cont2001,Mantegna1999,Mandelbrot1963,Mandelbrot1967}.
From the study of these time series, a set of statistical properties
common to many different markets, time periods and instruments, have
been identified. The universality of these properties is of interest
because the size, the participants and the events that affect the
changes of price (returns) in a certain market may differ enormously
from those that affect another. Yet, these investigations show that
the variations in prices indeed share non trivial statistical
properties, generically called \textit{stylized facts}. In this work
we present and study a model of a financial market and its
participants which reproduces these stylized facts.


The majority of approaches used today to model financial markets fall
into one of two categories: statistical models adjusted to fit the
history of past prices and Dynamic Stochastic General Equilibrium
(DSGE) models. The first kind of models are able to produce reasonable
representations and volatility forecasts of financial
systems\cite{Brownlees2011} as long as the statistical properties of
the prices with which they were calibrated do not change by a large
margin. The second kind of models assume a "representative agent'' for
each of the participant sectors in the financial system, each of these
agents attempting to their utility\cite{Sbordone2010}. To avoid
creating deterministic dynamics without periods of depression or
growth, DSGE models use exogenous stochastic terms which are supposed
to mimic the varying conditions of the market, such as sudden peaks in
the demand of a certain financial instrument or changes in the pricing
of a commodity.

Despite of the fact that these models are capable of providing some
explanations of the phenomena observed in financial markets, the
premises over which they are built are crude approximations of
reality\cite{Farmer2009,DeGrauwe2010} and as a such they are not
always useful to gain insight into statistical phenomena as rich at that
observed in financial time series.\\

This situation has given rise to the exploration of financial systems
as "complex systems" \cite{Bonabeau2002}. That is, to consider
financial markets as something closer to what they actually are:
systems where great number of different components interact amongst
each other in a way that gives rise spontaneously to the observed
macroscopic statistical properties.

Among the models which approach financial markets as complex systems,
there is a particular kind called ``Agent Based Models'' which employ
a bottom-up approach and allow the modeler to trace back the emergence
of the macroscopic statistical properties of the system as a
consequence of the microscopic behavioral traits of its constituent
agents\cite{Tesfatsion2006}. Several Agent Based Models have been
created that are capable of reproducing stylized facts and provide
possible microscopic explanations of their origins. These models have
been constructed, in general, in one of two ways: models in which the
agents do not use a particular set of strategies, but rather
participate in the market in a random fashion, and models in which the
agents follow different specific strategies inspired in actual
strategies used by participants of real markets, as we do in this
work.  The first type of models usually make use of market trading
structures similar to those used in real markets, such as double
auction order books, and as a consequence, the price formation is
directly driven by the offers (to buy and sell) supplied by the agents
\cite{Preis2006,Farmer2005,Maslov2000,Challet2003,Thilo2012}.  The
latter type of models usually have prices adjusted in a stochastic
manner\cite{Lux2000,Cont2007,Bertella2014,Alfi2009}.  Thus, while
models with ``intelligent'' agents employing different strategies in
realistic market environments have been proposed before
\cite{Chiarella2002,Chiarella2009,Consiglio2005,Licalzi2003}, in this
work we aimed to model the behavior of market participants following
the rules of thumb employed by real life traders, while keeping the
model as simple as possible. In particular, we do not pay much
attention to microeconomical foundations of the behavior of the
agents, such as rationality and utility maximization. Specifically, in
our model, we consider two types of agent: technical and
fundamental. Technical agents in our model follow a ``Moving average
oscilator'' strategy \cite{Brock1992}, which is commonly used by real
technical traders. These traders also incur in profit taking if the
price of the asset exceeds a certain threshold. Heterogeneity among
technical agents is achieved by assigning different parameters
(``personalities'') to different subsets of the technical agent
population. On the other hand, the fundamental agents in our model
``choose'' a fundamental price, and change it according to the influx
of news as well as the distance to the positions of the rest of the
agents in the market. The fundamental prices chosen by these agents,
and their reaction to the incoming news, differ amongst agents, as
happens in real life. Trading in the model is done through an order
book.

Since the model is constructed trying to mimic behavioral patterns
followed by the participants in real financial markets, we
expect that, if these behaviors are succesfully captured, however
simplified they may be, the resulting price statistics should
reproduce the stylized facts observed empirically. Specifically, the
stylized facts on which we focus in this paper, are the following:

\emph{Absence of auto-correlations}: The auto-correlation function of
the returns $R(t)$ is essentially zero for any value of the lag
(except at very short time in which there is a negative correlation
``bounce'' \cite{Cont2001}). The absence of auto-correlations has been
used as support for the efficient market hypothesis \cite{Bera2015}
since it implies that it is impossible to incur in
arbitrage\cite{Cristelli2014}.

\emph{Volatility Clustering} Notwithstanding the absence of
auto-correlations in the ``raw returns'' series, some non linear
functions of returns do exhibit auto-correlations that remain positive
for relatively long times. This behavior arises from the fact that the
returns have a tendency to ``agglomerate in time'' in groups of
similar magnitude but unpredictable sign \cite{Mandelbrot1963}.
 
\emph{Heavy tailed distribution of returns} The distributions of
price changes in real financial time series do not have a normal
distribution \cite{Mandelbrot1963,Bouchaud2001,Plerou1999}. Instead,
the distribution is characterized by having large positive values of
the kurtosis (for instance, the kurtosis for the Standard \& Poor's
index measured over time intervals of 5 minutes has been reported to
have a value of $\kappa \approx 16$\cite{Cont1997}).  Further, studies
of the complementary cumulative distribution of returns have shown
that it behaves approximately as a power law with an exponent $\beta
\in [2,4]$ \cite{Cristelli2014,Bouchaud2001}.

\emph{Asymmetry in the distribution of returns} In addition to being
heavy tailed, it has observed that in many markets, large negative
returns are more frequent than large positive returns. This asymmetry
is behind the negative skewness in the returns distribution which has
been reported in empirical studies\cite{Cont2001}.

\emph{Log-normal distribution of volatilities} The probability
distribution of the volatility of individual firm shares and of
indexes, defined as the average of the absolute returns over a time
window, is well approximated by a log-normal distribution in its
central part, while its tail is well adjusted by a power law with
exponent $ \mu \approx 3$\cite{Yanhui1999}.

In the next section we present a detailed discussion of the agent
based model we propose. The paper continues with a section in which we
present the results obtained in simulations of the model and we focus
on the stylized facts listed above, comparing the behavior of the
model with representative empirical data. We also study the effect of
varying the relative populations of agents as well as the parameters
that control the practice of profit taking by the technical agents in
the system. We end with a section of concluding remarks and
perspectives.

\section{Model}
\subsection{General Aspects}
The model represents a financial market in which $N$ agents trade a
single asset through a double auction order book in which the standing
orders are registered until executed.  In the model we only consider
market and limit orders \cite{Harris2002} of unit volume.

Like in actual financial markets, in the model, the population of
agents is divided into two different sub-populations, with each
sub-population employing one of two basic trading strategies:
fundamental analysis -by which a ``fundamental price'' $p_{f}$ is
estimated, and then the traders attempt to take advantage of the
deviations between $p_{f}$ and its present trading price $P_{t}$-; or
technical analysis -by which the trader tries to identify and exploit
trends in the price time series-.

These two types of strategies are representative of the main
strategies used in real life trading and were first introduced in the
Lux-Marchesi (LM) model \cite{Lux1999}. The effects of these
strategies on the dynamics of the price are opposed: while fundamental
agents tend to stabilize the prices around the average value of their
fundamental prices, technical agents tend to create periods of violent
price changes.  

The parameters controlling the behavior of each agent are assigned at
the beginning of each simulation, and even if two agents belong to the
same group (fundamental or technical) the difference in the values of
their controlling parameters will generate different "personalities"
within each strategy.

We make time run in discrete units corresponding to simulation steps
and on each simulation step, each fundamental agent will engage in
trading with a probability $p_{active}$ while technical agents will be
active when they observe a favorable trend or when they can obtain a
high immediate profit, as will be explained later.

In our system, every agent is assigned unlimited credit, and short
selling is allowed. These two liberties are meant to ensure that an
agent is able to engage in trading whenever it becomes active, thus
providing the market with enough liquidity.

Although the model we propose includes the main components of the
Lux-Marchesi model, there are important differences in the way in
which we designed both the agents and the market environment. Of
central importance is the fact that in our model the process of price
formation is directly governed by the demand and supply provided by
the agents and all the transactions are mediated through an order
book. Another important difference is the fact that by assigning
different parameters, we include heterogeneity within each
strategy. Further, in our model the only sources of ``exogenous''
randomness are the entry times of the fundamental agents; and the time
of arrival and nature of the news in the system, which in turn ellicit
randomly distributed reactions from fundamental agents.
 
\subsection{Types of Agents}

\subsubsection{Technical Agents}
As mentioned above, technical agents employ ``technical analysis'' in
an attempt to predict the future behavior of the price time series
with the purpose of exploiting the knowledge of that future behavior.

\begin{figure}[!htbp]
\centering
\includegraphics[width=1.0\textwidth]{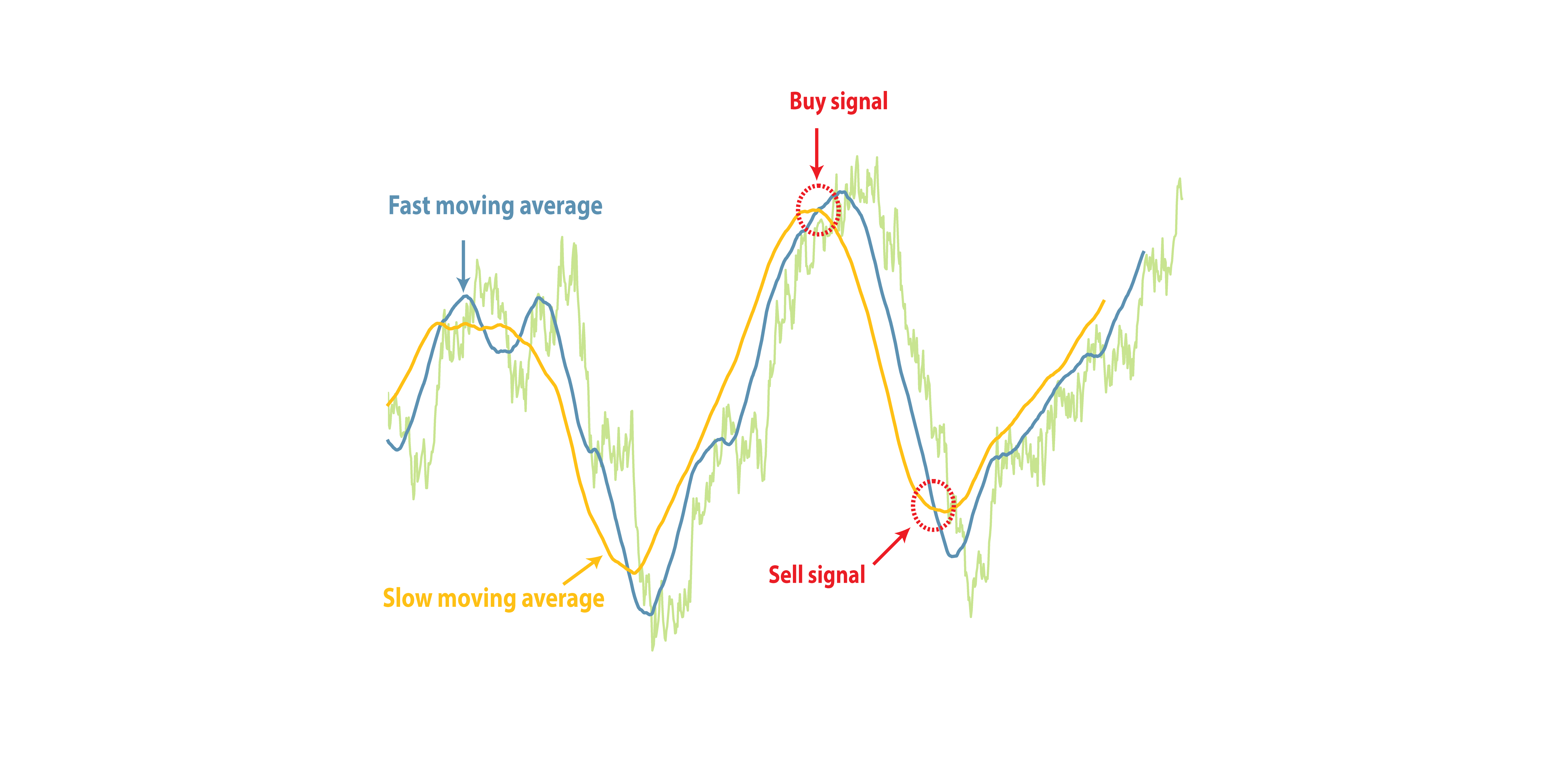}
\caption{Moving average-oscillator (MAO). This is a common technical
  indicator which is formed by two moving averages of different window
  sizes that are constantly observed. The moving average with the
  largest window size is called \textit{fast moving average} and the
  one with the smallest window is called \textit{slow moving
    average}. When the fast moving average crosses the slow one from
  below, a signal to buy is generated; conversely, when the fast
  moving average crosses the slow from above, a signal to sell is
  generated.}
\label{fig:MAO}
\end{figure}

In our model, technical agents utilize a technique used in real life
called Moving Average-Oscillator (MAO) \cite{Brock1992}, which
consists of a pair of moving averages with different window sizes: a
long period average called the \textit{slow moving average}, and a
short period average aptly called the \textit{fast moving
  average}. The fast moving average is intended to capture the
tendency of the price movements in a short term while the slow moving
average has the purpose of capturing the long term trend. Figure
\ref{fig:MAO} shows an example of this technical indicator.

When the fast average crosses the slow one from above, the MAO strategy
suggests that this is a ``signal to sell'', since the prices show a
short term tendency to fall below the long term trend captured by the
slow moving average.  Similarly, a ``signal to buy'' occurs when the
fast moving average crosses the slow one from below, since this can be
interpreted as the prices having a short time tendency to rise above
the long term trend.

We employ the MAO indicator in our model because while it is very simple and
easy to implement, it is representative of the plethora of technical
analysis tools and it is widely used in real markets\cite{LeBaron1992}.

In our model we use MAO indicators that differ in the window sizes of
the two averages which compose them. For each of these indicators
there is a population of technical agents following its evolution over
time and engaging in trading as a result of the signals that the
indicator generates. Further, when an indicator generates a signal to
either a buy or sell, each technical agent following that particular
indicator waits a particular time $t_{wait}$ before entering the action
suggested by the signal. This waiting time between the moment in which
the signal is generated and the moment in which an agent enters its
order is meant to allow the price time series to move in the direction
predicted by the indicator. If the agents were to immediately enter
their orders after they received a signal, they would not take
advantage of the rise or fall in prices that the trends point to. The
waiting time $t_{wait}$ of each technical agent is drawn from a
uniform distribution in the interval $[0,t_{max}]$, and assigned to
each agent from the beginning of the simulation.

A consequence of the way in which the MAO indicator is constructed, is
that the technical agents should have perfectly alternating order
flows, with a sell order following a previously entered buy order and
vice-versa. This alternation arises from the fact that the MAO
indicator generates signals when the two moving averages cross each
other and for any of the two directions of crossing: the fast average
crossing the slow one from below or from above, the next direction
will be necessarily of the opposite kind.

\begin{figure}[!htbp]
\centering
\includegraphics[width=0.8\textwidth]{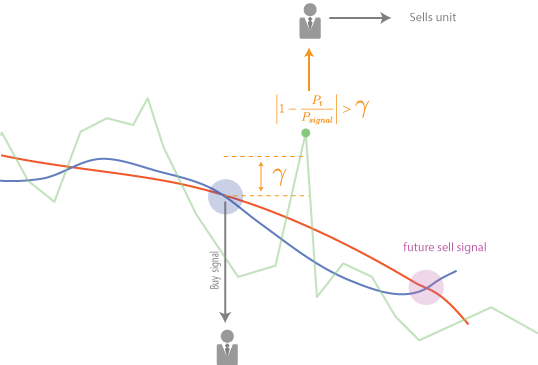}
\caption{Profit taking mechanism: If after observing a signal to buy,
  the prices rise enough (in our case this is defined as the moment at
  which $\left| 1-\frac{P_{t}}{P_{signal}} \right|$ exceeds a
  parameter $\gamma$), the technical agent will proceed to enter a
  sell market order. This practice is commonly used by traders to
  insure an immediate profit.}
\label{fig:ProfitTaking}
\end{figure}

There is, however, another mechanism which compels a technical agent
to engage in trading, aside from following the technical indicator.
This mechanism is \textit{profit taking} and it basically
consists in selling the asset when the price is sufficiently
high with respect to the price at which the last unit was bought,
irrespective of whether the MAO indicator generates a sell signal or
not, thus providing the agent with an immediate profit.  This is
implemented as follows, when a technical agent enters an order to the
book while following the indicator, that agent registers the price at
which the order was executed in a variable called $P_{signal}$. If the
price of the asset $P_{t}$ deviates from $P_{signal}$ by more than a
factor $\gamma$, the agent will proceed to enter a new sell order;
i.e. if after following a buy signal and entering the corresponding
buy order to the order book the price of the asset is greater than
$(1+\gamma)P_{signal}$, then the agent will place a sell market order,
securing in this way an immediate profit. Figure
\ref{fig:ProfitTaking} shows how profit taking is carried out in our
model.

The profit taking mechanism is introduced in our model because it is a
common practice in real financial markets and, as we will see, it
turns out to have a strong effect on the return statistics.

\subsubsection{Fundamental Agents}
A fundamental analysis trading strategy is based on two basic premises: the
first one being that every asset has an intrinsic ``fundamental
price'' $p_{f}$, and the second one, that in the short run, this
fundamental price may be incorrectly estimated by the market
participants but that in the long run, the market will correctly value
the asset and its price will eventually reach the fundamental price
$p_{f}$. An agent following a strategy of this kind will therefore buy
an asset when the price at which it is being traded is below his
estimation of its fundamental price $p_{f}$ and will sell the asset
when its price is above $p_{f}$. In this way a person following a
fundamental strategy will take advantage of the differences between
the prices at which the asset is traded over time and the fundamental
price; until the asset finally reaches its fundamental price.

When a fundamental agent becomes active, there are three available
actions that this agent can engage in: either to buy a unit of the
asset, to sell it (even short sell) or to abstain from either.  The
decision of whether to buy, sell or abstain from participating will
depend on the position of the agent's fundamental price $p_{f}$
relative to the price of the nearest best order (best ask or best
bid).

If $p_{f} > B_{sell}$, where $B_{sell}$ is the price of the best ask,
the agent will proceed to buy since there are agents willing to sell
for less than what the agent considers to be the correct
price. Similarly if $p_{f} < B_{buy}$, where $B_{buy}$ is the price of
the best buy, the agent will proceed to sell since there are agents
willing to buy offering more than the correct price.  If neither of
these two conditions is fulfilled, i.e. if $B_{sell} > p_{f} >
B_{buy}$ then there will be no competitive offers, since the lowest
price at which the agent could buy a unit of the asset is higher than
$p_{f}$, and the highest price at which it could sell a unit is lower
than $p_{f}$. Thus, when this condition arises the agent will abstain
from participating in the market.

\begin{figure}[!htbp]
\centering
\includegraphics[width=0.8\textwidth]{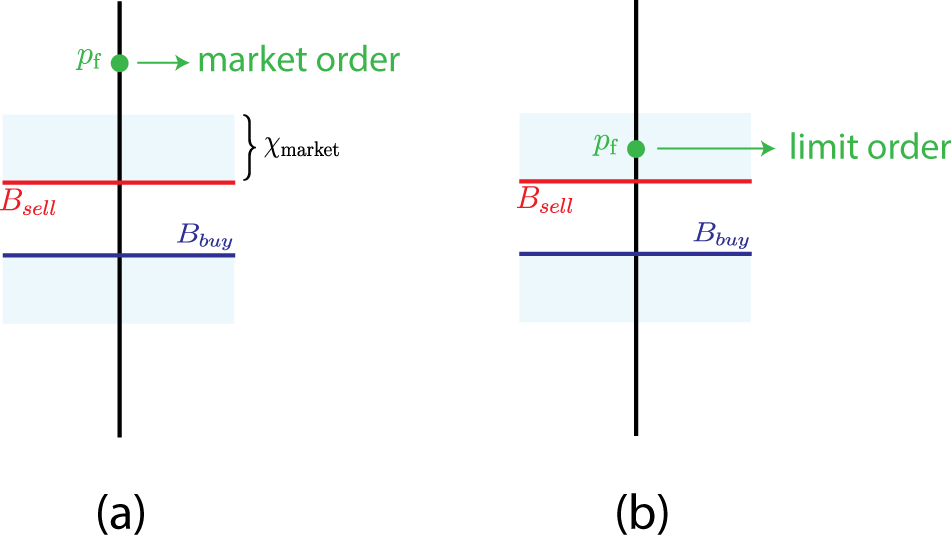}
\caption{Order selection algorithm for fundamental agents. In (a) we
  show the conditions that lead to a fundamental agent to introduce a
  market order: if the fundamental price $p_{f}$ is higher by more
  than a threshold $\chi_{market}$ (specific to each trader) with
  respect to the price of the nearest best order, the agent will
  proceed to enter a market order. Otherwise, the agent will proceed
  to enter a limit order (b). In the figure the orders would be
  ``buy'' orders as the agent's fundamental price lies above the best
  ask.}
\label{fig:FundsOrderSelection}
\end{figure}

When an agent decides to buy or sell, the decision to do so by
entering a limit or a market order will depend on the distance between
$p_{f}$ and the price of the nearest best order. Specifically, if the
agent decides to buy, it will do so by emitting a market buy order
when its fundamental price is above the price of the best sell offer
by more than a certain threshold $\chi_{market}$, i.e. when $p_{f} >
B_{sell}(1 + \chi_{market})$, and it will emit a limit buy order when
$p_{f}$ is below this threshold. Similarly, when the agent decides to
sell, it will do so by emitting a market sell order if its $p_{f}$ is
below the best buy offer by more than the threshold $\chi_{market}$,
i.e. when $p_{f} < B_{sell}(1 - \chi_{market})$, otherwise it will
emit a limit sell order.  Just like every other parameter defining the
behavior of a fundamental agents, every agent is assigned an
individual threshold $\chi_{market}$ from the beginning. The figure
\ref{fig:FundsOrderSelection} shows this decision making algorithm.

On the occasions in which a fundamental agent decides to enter a limit
order, the actual price of the order is extracted from a shifted symmetric
exponential distribution of the form:
\begin{displaymath}
      f(x;\lambda_{limit},\mu_{spread}) = \lambda_{limit}{}e^{- \left| \lambda_{limit}(x-\mu_{spread}) \right|}
\end{displaymath}
where $\mu_{spread}$ is the average price of the best orders:
$\mu_{spread} = \frac{1}{2}(B_{sell} + B_{buy})$. By assigning the
prices of limit orders in this way, they will have a greater tendency
to cluster around $\mu_{spread}$ which is a representative measure of
the central price at which the market participants are valuing the
asset. This behavior is intended to reflect the situation in which the
prices are not good enough to enter a market order, so the fundamental
agents will proceed to bargain with limit orders at prices that will
be close to the central price in the market.

\begin{figure}[!htbp]
\centering
\includegraphics[width=0.4\textwidth]{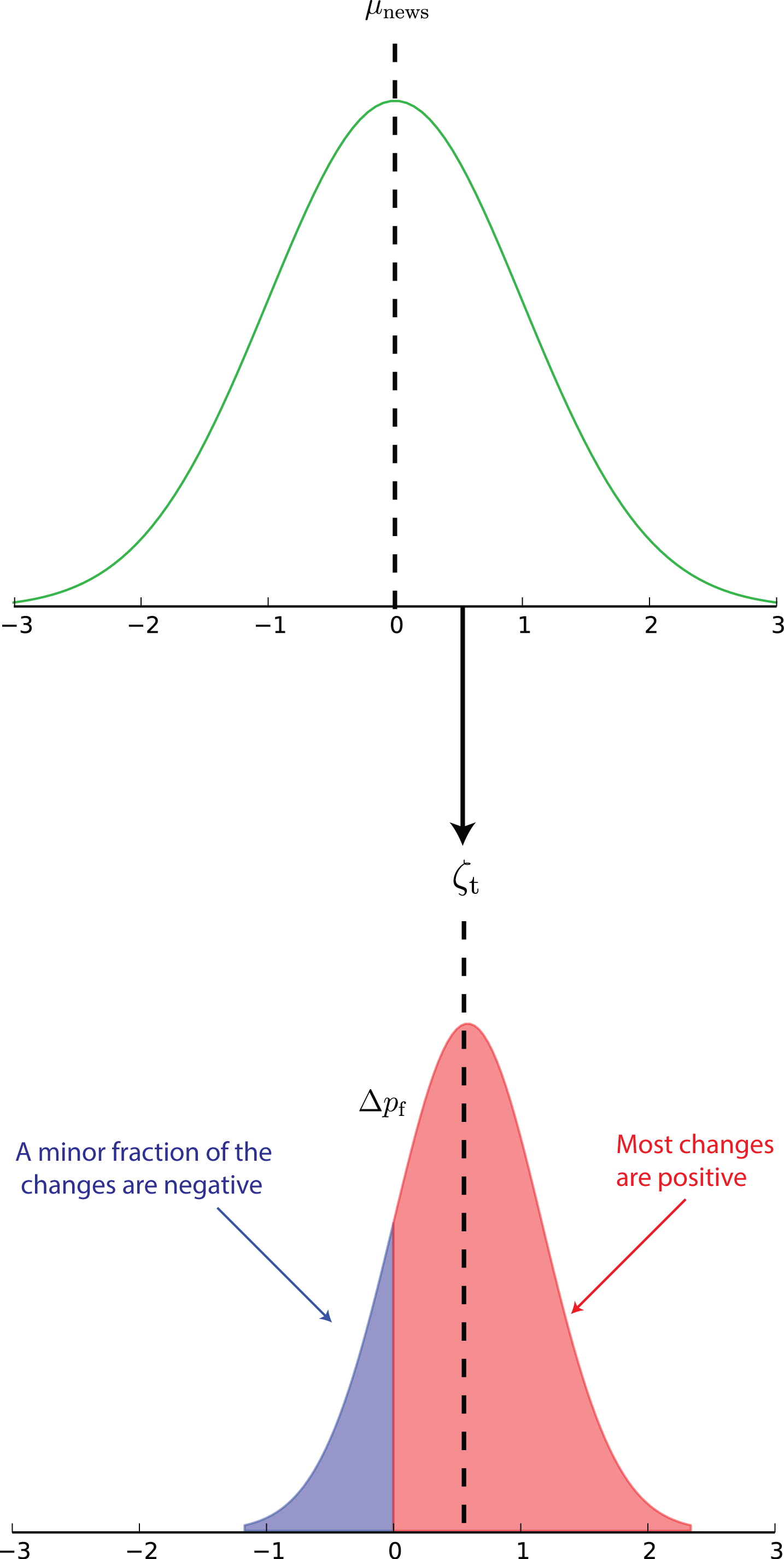}
\caption{News and their effects on fundamental prices. We model news
  as a sequence of IID Gaussian random variables. When a realization
  of this sequence, representing news being issued, occurs, the
  fundamental prices of each agent are adjusted from $p_{f}$ to $p_{f}
  + \Delta_{p_{f}}$ with $\Delta_{p_{f}}$ extracted from another
  normal distribution whose mean is equal to the value of the current
  news. In this way when highly positive news arrive, the majority of
  fundamental price changes will be positive; conversely, when highly
  negative news arrive, most price changes will also be negative.}
\label{fig:NewsEfectsFunds}
\end{figure}

In real life, $p_{f}$ is determined by each fundamental trader, and
then adjusted as time goes by, according to the appearance of news
concerning the well being of whatever underlies the asset. To include
this feature of fundamental analysis in our model, we introduce a flow
of news modeled as a sequence of IID random variables $\zeta_{t}$
taken from a normal distribution with mean $\mu_{news}$ and variance
$\sigma_{news}$. The time intervals betwen succesive news are taken
from a Poisson distribution. Here, $\zeta_{t}$ represents the mean
value by which the news will change the fundamental prices of the
asset. When, in the context of our model, news are issued at a given
time $t$, each fundamental agent adjusts its fundamental price from
$p_f(t)$ to $p_{f}(t) + \Delta{}p_{f}(t)$ where $\Delta{}p_{f}(t)$ is
again extracted from a normal distribution with mean $\zeta_{t}$ and
variance $\sigma_{\Delta{}p_{f}}$ as illustrated in Figure
\ref{fig:NewsEfectsFunds}. Thus, the majority of fundamental agents
will change their prices accordingly with the sign of $\zeta_{t}$,
however, depending on the magnitude of the news, some agents may even
extract a $\Delta{}p_{f}$ with an opposite sign to $\zeta_{t}$. This
diversity of response to a news item attempts to reflect the
possibility of diverse interpretations of the information by the
fundamental agents.  The fundamental price of each agent is chosen
from a uniform distribution at the beginning of a simulation.

Finally, although a fundamental agent bases its trading strategy in
the differences between its fundamental price and the prices at which
the market values the asset, if too large a difference is present, the
agent will try to get closer to the central market price
$\mu_{spread}$. This feature is meant to capture the attention that a
fundamental agent pays to the opinions of the whole population of
agents, which constitutes a mild manner of ``herding behavior''. If
the valuation of the fundamental price that an agent has is too far
from the price at which it is being traded, the agent will move its
fundamental price closer to the central price $\mu_{spread}$. This can
be interpreted as a precautionary move by the agent since such a big
difference between $p_{f}$ and $\mu_{spread}$ could point to
information that was not incorporated in the determination of his
fundamental price, or that an ineffective incorporation of the
available information was made.

To determine when the difference between $p_{f}$ and $\mu_{spread}$ is
``too big'', each agent compares this difference with a threshold
$\chi_{opinion}$, if at the time a fundamental agent becomes active,
such agent observes that

\begin{equation*}
	\chi_{opinion} < \left| 1 - \frac{p_{f}}{\mu_{s_{spread}}} \right| 
\end{equation*}

Then the agent will adjust its price to get closer to $\mu_{spread}$ in the following way:
\[ p_f = \left\{
\begin{array}{ll}
    \mu_{spread}\left(1+\chi_{opinion } \right),  & if  \,\, p_{f} \geq\mu_{spread} \\
     \mu_{spread}(1-\chi_{opinion}), & if  \,\ p_{f} < \mu_{spread}
\end{array} 
\right. \]

Thus, the agent will get as close to $\mu_{spread}$ as the
maximum tolerance ($\chi_{opinion}$) between its opinion and the
opinion of the population ($\mu_{spread}$) allows.


\section{Results}

In this section we present the results obtained in various
simulations. Although these results correspond to a particular set of
values for the parameters, reasonable changes in the values of these
parameters generate the same qualitative properties in the statistics
of the model. It is of critical importance for the stability of the system to have a
flow of limit orders (liquidity) capable of filling the gaps that are
created when market orders enter the order book. To achieve this, the
parameters that govern the flow of limit and market orders emitted by
the agents must not give rise to bursts of market orders with a volume
so large that one side of the order book is emptied. It is in this sense
that we speak above of reasonable changes in the values of the
parameters. Thus, for example, if we were to allow greater volumes of
market orders to be placed within shorter time windows, say, by
including a larger number of technical agents in a simulation, then,
the parameters that affect the input of limit orders must be chosen
accordingly, in such a way that the fundamental agents have enough
time to restore the liquidity consumed by the increased number of
market orders.

Unless otherwise stated, the following results were obtained with a population of 1000
fundamental agents and 1500 technical agents divided into two groups
of 750 agents with technical indicators made of moving averages with
window sizes of 4000 and 2000 time steps for one group and 2000 and
1000 time steps for the other. The other parameter values used for
this run are shown in Table \ref{tab:paramValues_1}.

\begin{table}[!htbp]
\centering
	\begin{tabular}{ccc}
		\hline
		\cline{1-2}
		 Parameter & Value \\
		 \hline
	     $P_{active}$ &  0.15\\
         $p_{f}$(initial) & [20.0, 25.0]\\
         $\chi_{market}$ & [0.005, 0.25]\\
	     $\chi_{opinion}$ & [0.01, 0.1]\\
         $\sigma_{\Delta_{p_{f}}}$ & 0.2\\
		 $\lambda_{limit}$ & 3\\
         $\mu_{news}$ & 0\\	
         $\sigma_{news}$ & 0.1\\	
         $f_{news}$ & 100\\	
         $\gamma$ & 0.01\\	
         $t_{wait}$ & [0, 50]\\	
		\hline
	\end{tabular}
    \caption{Values of the parameters corresponding to the results
      presented in this paper (ranges indicate that the parameters for
      each agent were taken from a uniform distribution in within the
      specified values).}
    \label{tab:paramValues_1}
\end{table}


As is frequently the case for many financial models, some of the
parameters defined in our model may not have a clear connection to
observables in real life, and even when observables similar to the
parameters in our model exist, attempting to estimate their values is
somewhat ambiguous. Thus, we chose values which allowed the
simulations to run in a stable manner and that generated statistical
properties similar to those observed in real markets.  Interestingly,
the model is rather robust and produces similar relevant results for a
wide range of parameter values.  The values of the parameters we
employed for the results we present below are therefore, just an
election among many different elections we made within the range of
useful parameter values.

\begin{figure}[!htbp]
\centering
\includegraphics[width=0.7\textwidth]{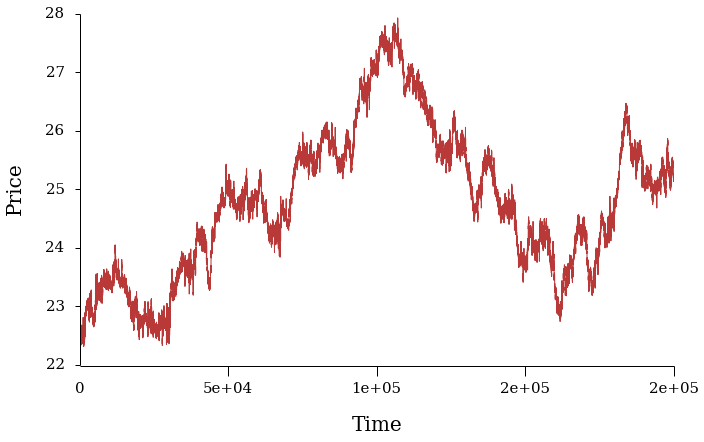}
\caption{Representative time series of asset prices, determined as the
  last price the asset was traded at each time step (``closing price''). }
\label{fig:PriceTimeSeries_FullModel}
\end{figure}

\begin{figure}[!htbp]
    \centering
    
    \begin{subfigure}[b]{0.7\textwidth}
        \includegraphics[width=\textwidth]{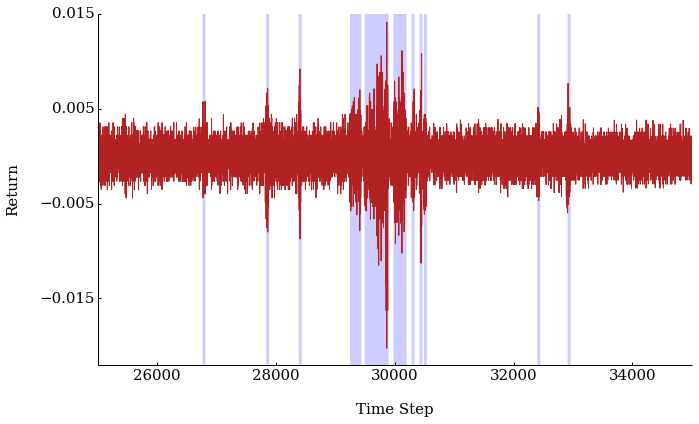}
        \caption{Returns corresponding to the simulation.}
        \label{fig:ReturnsTimeSeries_FullModel}
    \end{subfigure}

    \begin{subfigure}[b]{0.7\textwidth}
        \includegraphics[width=\textwidth]{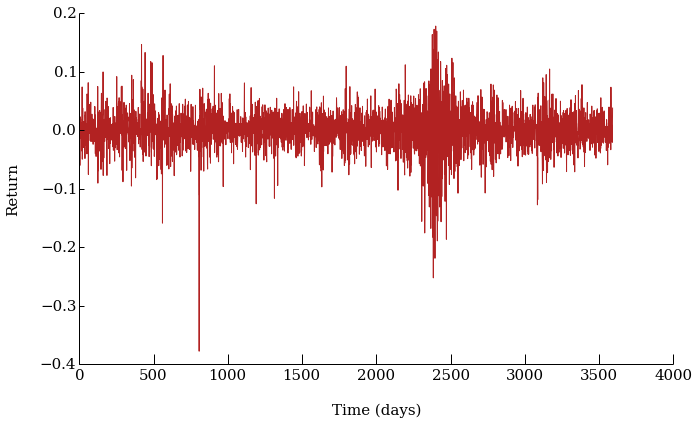}
        \caption{Returns from Consol Energy Inc. Data obtained from QuantQuote\cite{QuantQuote}.}
        \label{fig:ReturnsTimeSeries_Real}
    \end{subfigure}

    \caption{Returns time series for the simulation with a time lag
      $\tau=1$ (a) and comparison with empirical data from Consol
      Energy Inc (b). The blue shaded regions show the times in which
      technical agents were active, as can be seen, these times
      coincide with the periods with the largest changes of
      price.}\label{fig:VolatiulityClusterings}
\end{figure}

We begin by showing the time series corresponding to the prices and
logarithmic returns, defined as $ r(t) = \log(P_t/P_{t-\tau})$, for a
given lag $\tau$, generated by our model. These are shown in figures
\ref{fig:PriceTimeSeries_FullModel} and
\ref{fig:ReturnsTimeSeries_FullModel} respectively. The blue bars in
figure \ref{fig:ReturnsTimeSeries_FullModel} signal the time steps in
which technical agents were active. The bursts of greater volatility
coincide with the activity of the technical agents while the times in
which only fundamental agents were active (trading) present lower
volatility.

\begin{figure}[!htbp]
\centering
\includegraphics[width=0.7\textwidth]{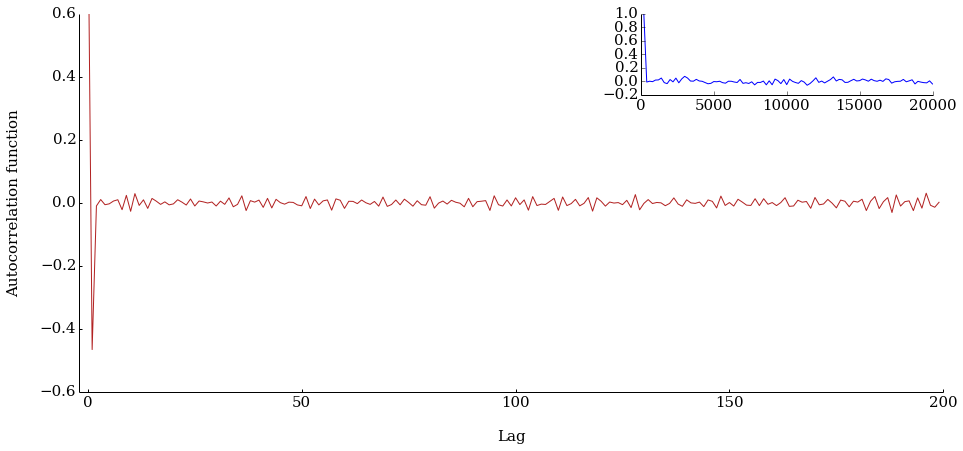}
\caption{Auto-correlation functions of returns. There are essentially
  no correlations for any value of the lag, except for a negative
  correlation that lasts for a few steps at the beginning. This
  phenomenon is also observed in real returns series and has been
  called \textit{bid-ask bounce}\cite{Cont2001}. The main figure corresponds to the
  autocorrelation function of the returns calculated every time step
  and the inset figure to the returns calculated every 50 steps.}
\label{fig:RawReturnsACF_FullModel}
\end{figure}

In figure \ref{fig:RawReturnsACF_FullModel} we show the
auto-correlation function of the returns, the blue line corresponds to
the returns calculated time step by time step. In the inset we show
the auto-correlation function for returns calculated every 50 steps,
in both cases it can be seen that the auto-correlation is essentially
zero for any value of the lag. It is interesting to note that the
phenomenon know as ``bid-ask bounce'' can be observed in the returns
generated by our simulations. This phenomenon consists in the presence
of negative values of the auto-correlation function at very short lags
and it is attributed to the fact that most transactions take place
near the best ask or best bid and tend to bounce between these two
values\cite{Cont2001}.

\begin{figure}[!htbp]
    \centering
    
    \begin{subfigure}[b]{0.7\textwidth}
        \includegraphics[width=\textwidth]{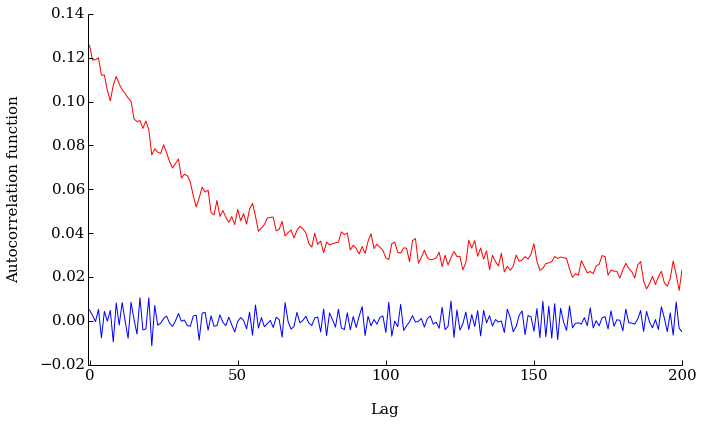}
        \caption{Direct (blue) and absolute (red) auto-correlation
          functions of returns from our the simulation.}
        \label{fig:VolatilityClusteringACF_FullModel}
    \end{subfigure}

    \begin{subfigure}[b]{0.7\textwidth}
        \includegraphics[width=\textwidth]{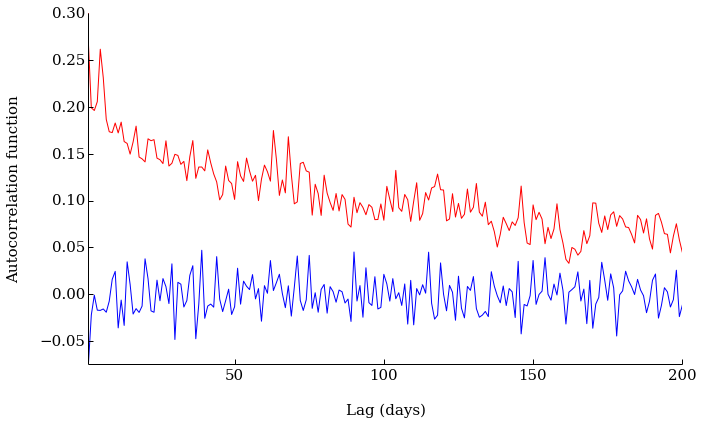}
        \caption{Corresponding auto-correlation functions from Airgas
          Inc. Data obtained from QuantQuote\cite{QuantQuote}.}
        \label{fig:AutocorrFuncts_Real}
    \end{subfigure}

    \caption{Returns auto-correlation function for the simulation (a)
      and comparison with empirical data from Airgas Inc (b). While
      the auto-correlation of the direct returns (blue lines) is zero, the
      auto-correlation of the absolute value of the returns (red lines)
      remains positive for a long period of time, and decays slowly to
      zero.}\label{fig:AutoCorrFuncts}
\end{figure}

In figure \ref{fig:VolatilityClusteringACF_FullModel} we present the
comparison between the auto-correlation of the returns (blue line) and
the auto-correlation of the absolute value of the returns (red
line). We observe that the auto-correlation function of the
absolute returns remains positive over a long time interval,
and that it decays slowly to zero. Figure
\ref{fig:AutocorrFuncts_Real} illustrates the same auto-correlation
functions for a representative company listed in the Standard \& Poor's 500.

\begin{figure}[!htbp]
    \centering
    ~ 
    \begin{subfigure}[b]{0.7\textwidth}
        \includegraphics[width=\textwidth]{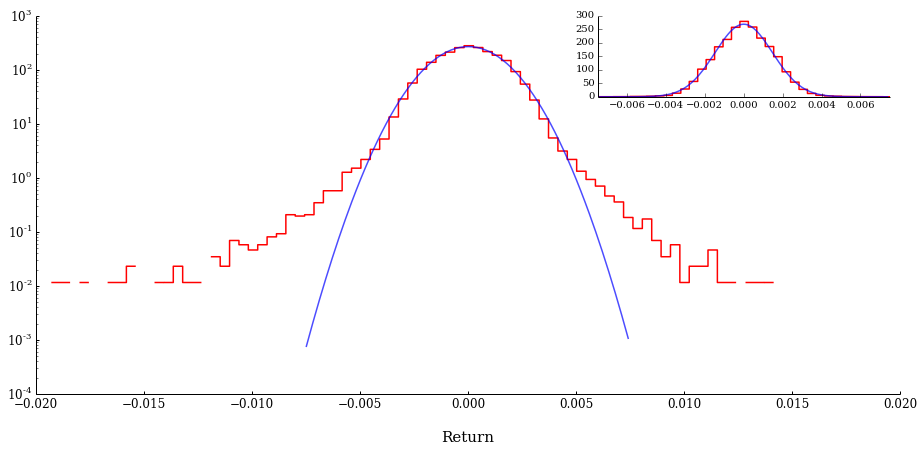}
        \caption{Returns PDF corresponding from the simulation.}
        \label{fig:ReturnsPDF_FullModel}
    \end{subfigure}

    \begin{subfigure}[b]{0.7\textwidth}
        \includegraphics[width=\textwidth]{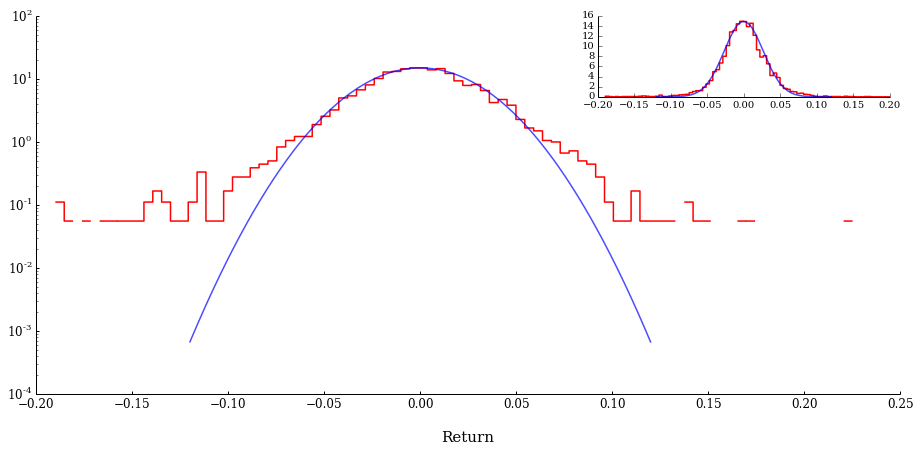}
        \caption{Returns PDF from United States Steel Corporation. Data obtained
          from QuantQuote\cite{QuantQuote}.}
        \label{fig:ReturnsPDF_Real}
    \end{subfigure}
    \caption{Returns PDF from the simulation (a) and comparison with
      empirical data from  United States Steel Corporation, Inc. The tails of the
      distribution (red line) are clearly heavier than those of a
      normal distribution (blue line). }\label{fig:ReturnsPDFs}
\end{figure}

Figure \ref{fig:ReturnsPDF_FullModel} shows the distribution function
of returns from our model. This distribution shows heavier tails than a
normal distribution with the same mean and standard deviation and it
is possible to observe that the left tail is heavier than the right
one. For comparison, figure \ref{fig:ReturnsPDF_Real} illustrates the
distribution function of returns for a representative company listed
in the Standard \& Poor's 500.

\begin{figure}[!htbp]
    \centering
    ~ 
    \begin{subfigure}[b]{0.7\textwidth}
        \includegraphics[width=\textwidth]{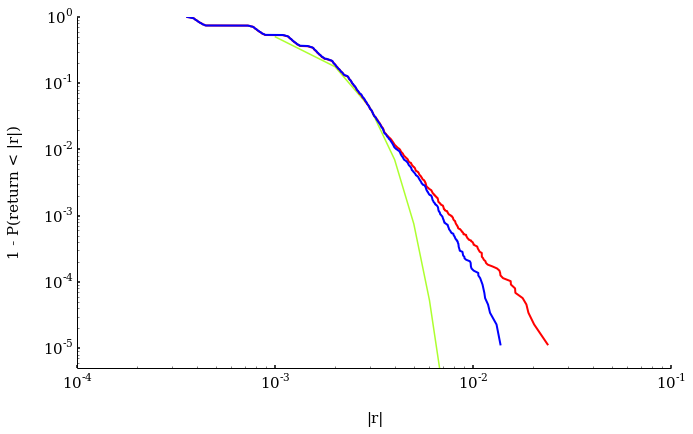}
        \caption{Returns CDF corresponding to the simulation.}
        \label{fig:ReturnsCDF_FullModel}
    \end{subfigure}

    \begin{subfigure}[b]{0.7\textwidth}
        \includegraphics[width=\textwidth]{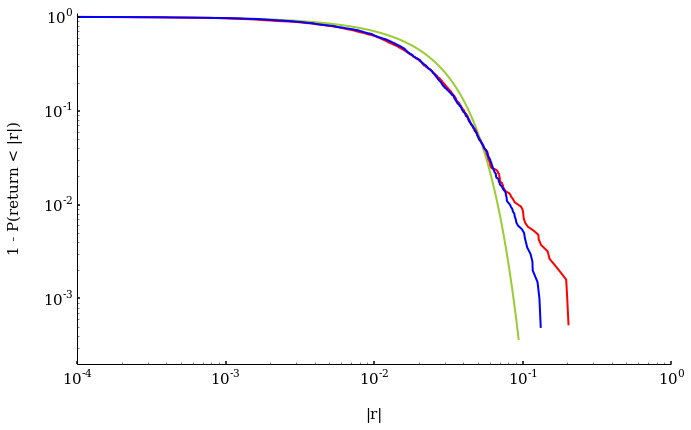}
        \caption{Returns CDF from Anadarko Petroleum Corp. Data
          obtained from QuantQuote\cite{QuantQuote}.}
        \label{fig:ReturnsCDF_Real}
    \end{subfigure}
    \caption{Comparison of the positive and negative returns CDF from
      the simulation (a) and from empirical data for Anadarko
      Petroleum Corp (b). It can be seen that the left tail of the
      distribution (red line), corresponding to the negative returns,
      is heavier than the right tail (blue line), corresponding to the
      positive returns. This is related to the negative skewness
      observed in the distribution.}
    \label{fig:CCDFs}
\end{figure}

\begin{figure}[!htbp]
  \begin{subfigure}[b]{0.5\textwidth}
    \includegraphics[width=\textwidth]{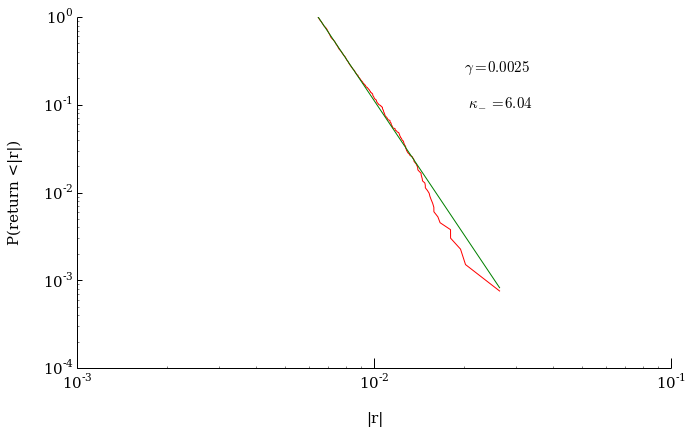}
    \caption{Negative returns}
    \label{fig:Left_PowerlawFit_0025}
  \end{subfigure}
  \hfill
  \begin{subfigure}[b]{0.5\textwidth}
    \includegraphics[width=\textwidth]{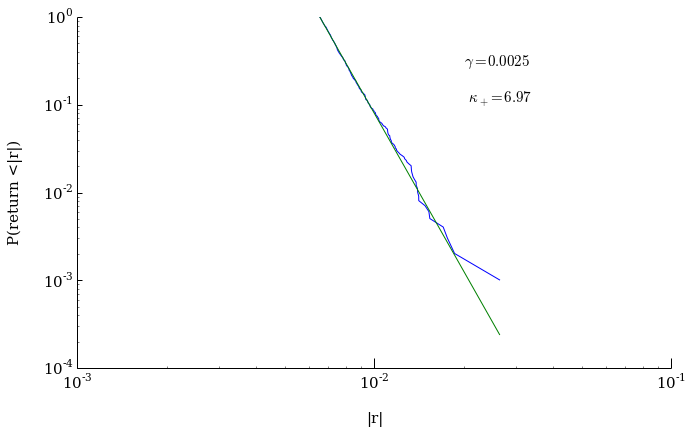}
    \caption{Positive returns}
    \label{fig:Right_PowerlawFit_0025}
  \end{subfigure}
  \caption{Complementary cumulative distribution functions and their
    powerlaw fits for $\gamma=0.0025$. Here $\kappa_-$ and $\kappa_+$
    are the exponents of the powerlaw fits for the left and right tail
    correspondingly.}
\end{figure}

\begin{figure}[!htbp]
  \begin{subfigure}[b]{0.5\textwidth}
    \includegraphics[width=\textwidth]{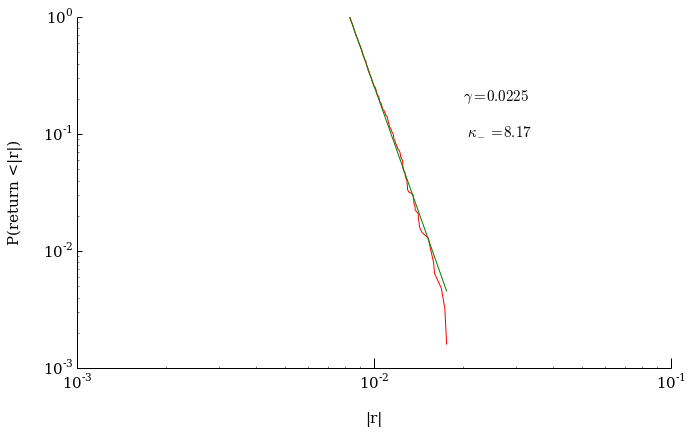}
    \caption{Negative returns}
    \label{fig:Left_PowerlawFit_0225}
  \end{subfigure}
  \hfill
  \begin{subfigure}[b]{0.5\textwidth}
    \includegraphics[width=\textwidth]{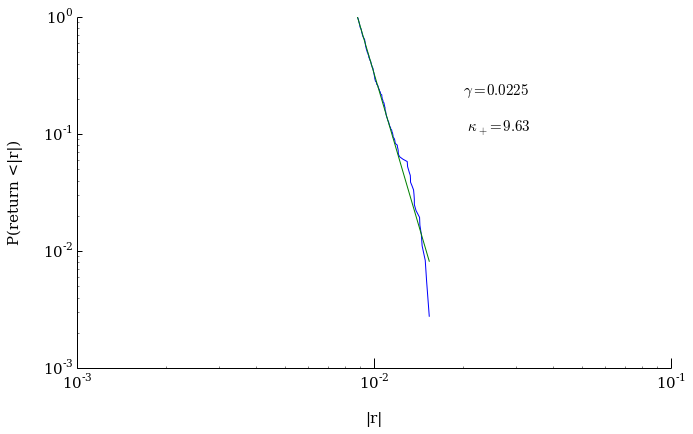}
    \caption{Positive returns}
    \label{fig:Right_PowerlawFit_0225}
  \end{subfigure}
  \caption{Complementary cumulative distribution functions and their powerlaw fits for $\gamma=0.0225$.}
\end{figure}

\begin{figure}[!htbp]
  \begin{subfigure}[b]{0.5\textwidth}
    \includegraphics[width=\textwidth]{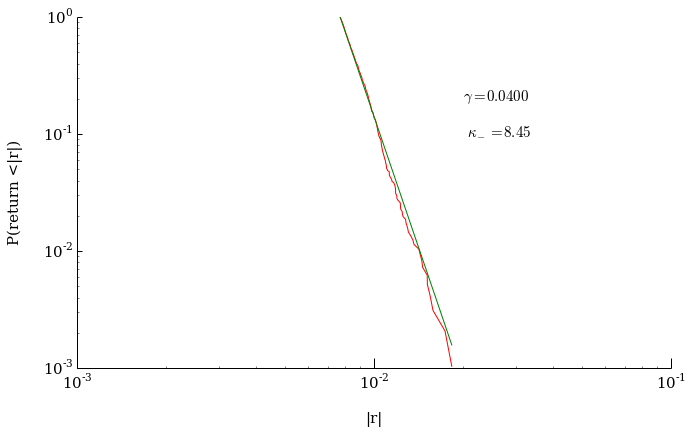}
    \caption{Negative returns}
    \label{fig:Left_PowerlawFit_0400}
  \end{subfigure}
  \hfill
  \begin{subfigure}[b]{0.5\textwidth}
    \includegraphics[width=\textwidth]{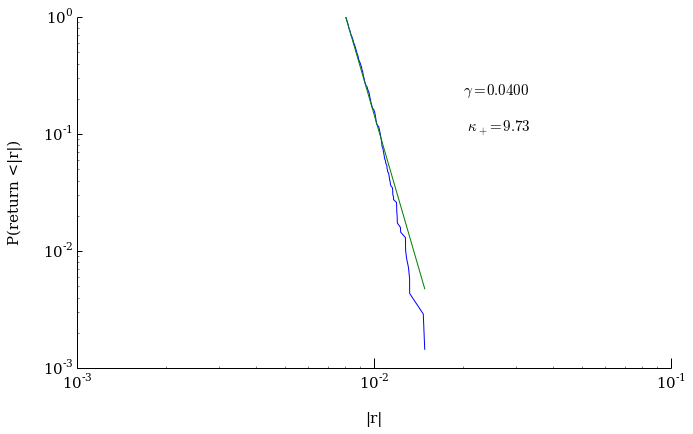}
    \caption{Positive returns}
    \label{fig:Right_PowerlawFit_0400}
  \end{subfigure}
  \caption{Complementary cumulative distribution functions and their powerlaw fits for $\gamma=0.0400$.}
\end{figure}

Figure \ref{fig:ReturnsCDF_FullModel} shows the cumulative
complementary distribution of positive and negative returns,
highlighting the asymmetry between losses and gains. The tail of the
distribution of negative price changes is significantly heavier than
the distribution of positive changes, a fact that is consistent with
the negative skewness displayed by the returns distribution.  Figure
\ref{fig:ReturnsCDF_Real} illustrates the corresponding distributions
for a representative company listed in the Standard \& Poor's 500.  In
addition to the asymmetry, it can be seen in figure
\ref{fig:ReturnsCDF_Real} that the tails of the distribution of returns seem to follow power law behavior.  To test how well a power law fits the data, we used the
python package ``powerlaw''\cite{Alstott2014}. Figures
\ref{fig:Left_PowerlawFit_0025},\ref{fig:Right_PowerlawFit_0025},
\ref{fig:Left_PowerlawFit_0225},\ref{fig:Right_PowerlawFit_0225},
\ref{fig:Left_PowerlawFit_0400},\ref{fig:Right_PowerlawFit_0400} show
fits for three different values of the parameter $\gamma$.  As can be
seen in the figures, both tails of the distribution are rather well
described by power laws.

The goodness of fit tests performed by the ``powerlaw'' package throw
as a result the log-likelihood ratio $R$ between two different
candidate distributions. In this test $R>0$ (respectively $R<0$) when
the first distribution is more (less) likely to describe the data than
the second distribution\cite{Clauset2009}.  To assess how much the sign of $R$ was
affected by the statistical fluctuations, the significance $p$, gives
the probability of measuring a given value of $R$ under the assumption
that its real value is close to zero. A small value of $p$ means that
it is unlikely that the measured value of $R$ is a product of the
fluctuations, and, as a consequence, that its sign can be trusted as
an indicator of which distribution provides a better fit for the
data. The average values of $R$ and $p$ for simulations with different
values of $\gamma$ are presented in table \ref{tab:Powerlaw_Fits}. For
each value of $\gamma$ in the table an ensemble of 50 simulations was
run and the mean values of the loglikelihood for the left tail
($<R_{-}>$) and right tail ($<R_{+}>$) as well as the significance
values $<p_{-}>$ and $<p_{+}>$ are presented.


\begin{table}[!htbp]
\centering
	\begin{tabular}{ccccc}
	\hline
	\cline{1-5}
		 & $<R_{-}>$  & $<p_{-}>$  & $<R_{+}>$ & $<p_{+}>$\\
		 \hline
	     $\gamma=0.0025$ & -0.006 & 0.595 & -0.032 & 0.057\\
	     $\gamma=0.0225$ & 0.004 & 0.597 & -0.053 & 0.623\\
	     $\gamma=0.0400$ & -0.050 & 0.609 & 0.203 & 0.489\\
   		\hline
	\end{tabular}
    \caption{Values of the mean log-likelihood ratios $<R>$ between the
      powerlaw and lognormal fits and of the mean significance values
      $<p>$. The values are presented for empirical data and for three
      representative cases of our model with different values of
      $\gamma$. Here $<R_{-}>$ and $<R_{+}>$ stand for the
      log-likelihoods of the left and right tails,
      correspondingly. Similarly, $<p_{-}>$ and $<p_{+}>$ stand for
      the mean significance values for the left and right tails.}
    \label{tab:Powerlaw_Fits}
\end{table}

\begin{table}[!htbp]
\centering
	\begin{tabular}{cccc}
	\hline
	\cline{1-4}
		 $<R_{-}>$  & $<p_{-}>$  & $<R_{+}>$ & $<p_{+}>$\\
		 \hline
	      0.258 & 0.399 & 0.403 & 0.338\\

   		\hline
	\end{tabular}
    \caption{Values of the mean log-likelihood ratios $<R>$ between the
      powerlaw and lognormal fits and of the mean significance values
      $<p>$ for the empirical data.}
    \label{tab:Powerlaw_Fits_Empirical}
\end{table}

Table \ref{tab:Powerlaw_Fits_Empirical} shows the mean loglikelihood
ratios and significance values measured in the empirical data. From
the empirical data set it can be seen that although the ratios point
to a power law as the best fit when compared to a lognormal
distribution, the significance values are again high enough ($> 0.10$) to
make inconclusive the test. Similarly, in the data set generated from
the simulations, the significance values are too high to ascertain
whether a powerlaw distribution is a better fit than a
lognormal. Nevertheless, the power law fits seem to be a very good
description of the behavior in both tails of the distribution for all
three cases of $\gamma$, which span the range from very frequent to
very scarce engagement in profit taking.

\begin{figure}[!htbp]
     \centering
     ~ 
     \begin{subfigure}[b]{0.7\textwidth}
         \includegraphics[width=\textwidth]{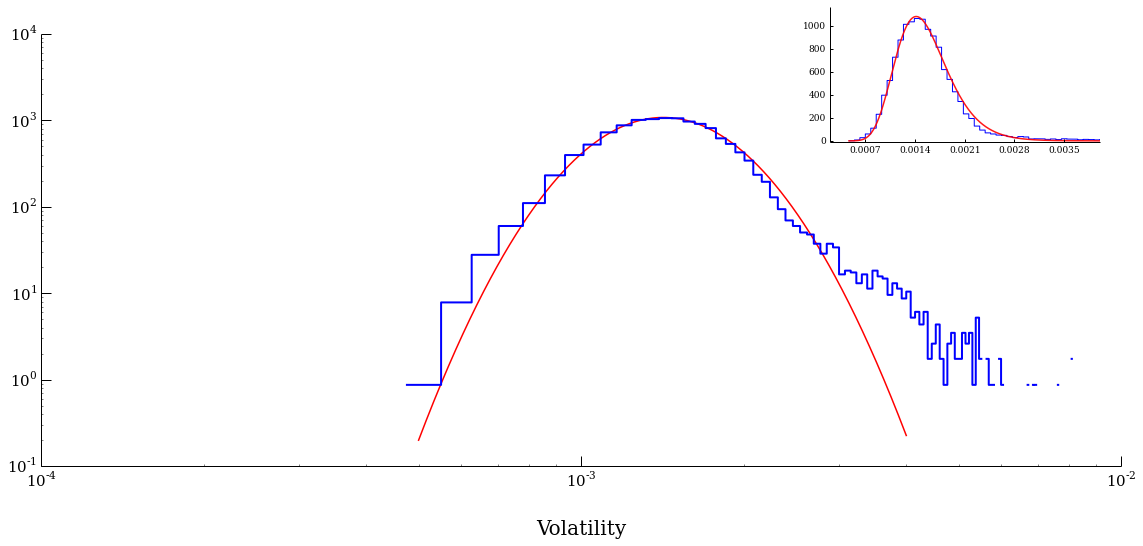}
         \caption{Volatilities PDF. It can be seen that, while the
           distribution of returns is not well described by a
           log-normal distribution, the central region is
           qualitatively similar to one, but the right tail is
           considerably heavier.}
         \label{fig:VolatilitiesPDF_FullModel}
     \end{subfigure}
  
     \begin{subfigure}[b]{0.7\textwidth}
         \includegraphics[width=\textwidth]{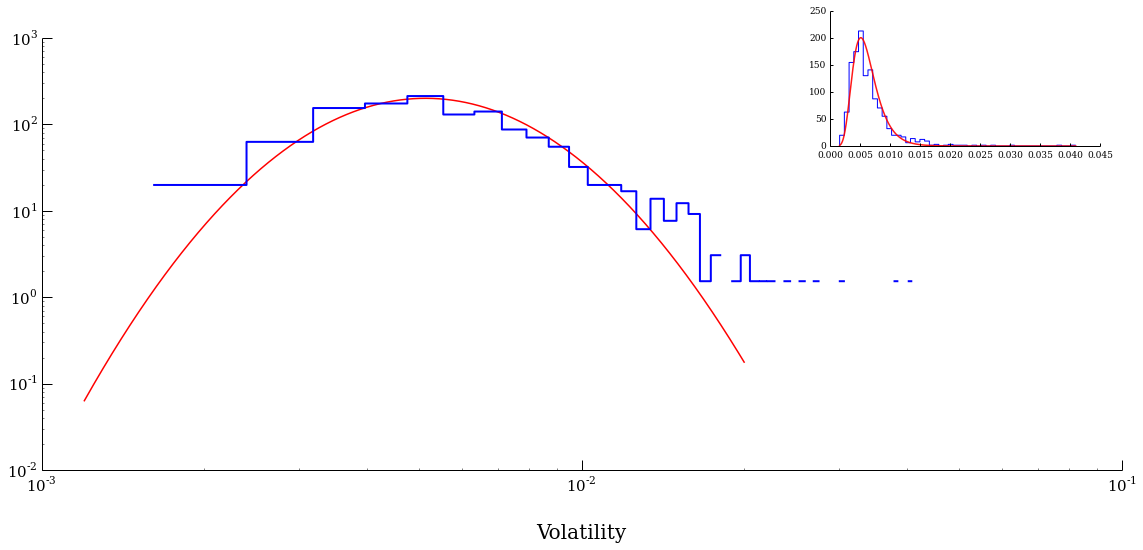}
         \caption{Volatility distribution of Exelon Corp. Data obtained
           from QuantQuote\cite{QuantQuote}.}
         \label{fig:VolatilitiesPDF_Real}
     \end{subfigure}
     \caption{Distribution of volatilities for a simulation with both
       fundamental agents and technical agents (a) and comparison with
       empirical data from the Standard \& Poor's 500 (b). Data
       obtained from Yahoo Finance.}\label{fig:LogNormals}
 \end{figure}

\begin{figure}[!htbp]
\centering
\includegraphics[width=0.7\textwidth]{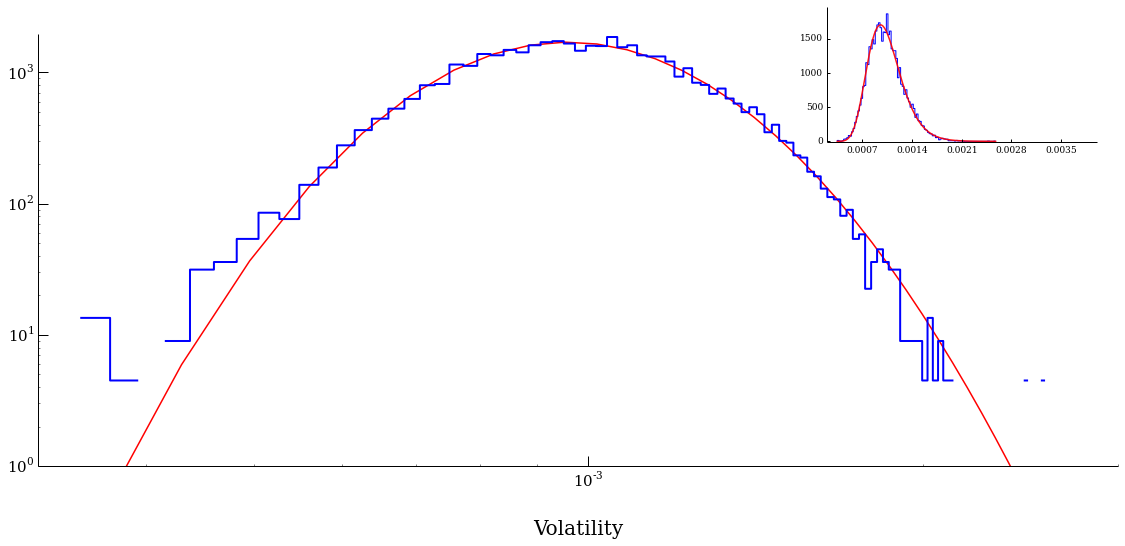}
\caption{Volatilities of a simulation without technical
agents. When only fundamental agents are used in a
simulation, a log-normal distribution is a remarkably good
description of the distribution of volatilities.}
\label{fig:VolatilitiesPDF_IncompleteModel}
\end{figure}

In figure \ref{fig:VolatilitiesPDF_FullModel} we present the
distribution of volatilities measured as the average of the absolute
value of returns $|r(t)|$ over a time window $T=n\Delta{}t$, i.e.
\begin{displaymath}
	V_{T}(t) = \frac{1}{n}\sum\limits_{t^{'}=t}^{t+n-1} \left| r(t^{'}) \right|
\end{displaymath}
For the present result we took values of $n=30$ and $\Delta{}t=1$ time
steps. The distribution of volatilities is not well described by a
log-normal distribution, however, the central part of the distribution
may be approximated by one\cite{Yanhui1999}. On the other hand, when
we remove the technical agents from the simulation, the volatilities
are remarkably well described by a log-normal distribution as shown in
Figure \ref{fig:VolatilitiesPDF_IncompleteModel}, which corresponds to
a run with the same parameter values described in Table
\ref{tab:paramValues_1} without technical agents.


To assess how well a lognormal distribution fits the volatilities, we
performed a Kolmogorov-Smirnov test on the empirical data and on four
different sets of data generated with our model. The p-values obtained
from these tests are presented in table \ref{tab:Lognormal_Fits}. 
Even in the case with $\gamma =0.0025$ which
generated data which clearly deviates from a lognormal distribution at
the tails, the average p-value is still high enough to make the
rejection of the lognormal hypothesis difficult. The values obtained with
the model are very similar to the value of the average p-value
measured from the empirical data, which is at $0.47$.

\begin{table}[!htbp]
\centering
	\begin{tabular}{ccccc}
	\hline
	\cline{1-5}
		 & $\gamma=0.0025$  & $\gamma=0.0150$  & $\gamma=0.0300$ & $\gamma=0.0400$\\
		 \hline
	     p-value & 0.21 & 0.42 & 0.49 & 0.50\\
   		\hline
	\end{tabular}
    \caption{Values of the mean p values corresponding to the goodness
      of fit of a lognormal distribution for the distribution of
      volatilities for four representative cases of our model with
      different values of $\gamma$.}
    \label{tab:Lognormal_Fits}
\end{table}

\begin{figure}[!htbp]
    \centering

    \begin{subfigure}[b]{0.7\textwidth}
        \includegraphics[width=\textwidth]{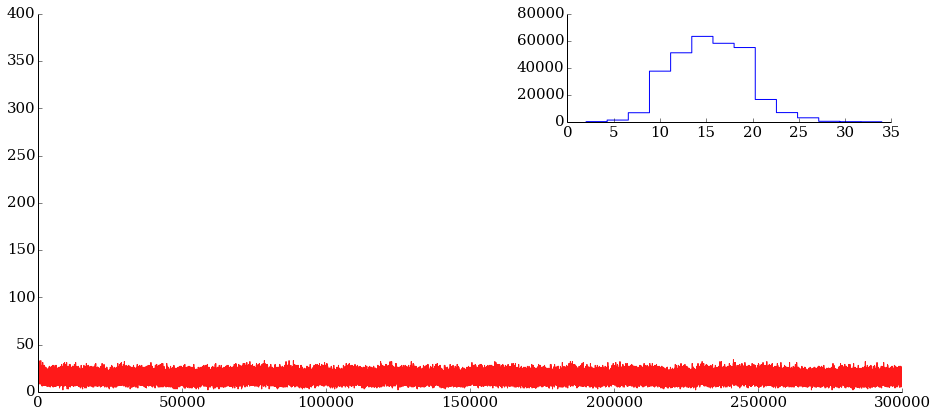}
        \caption{Trading volumes for a simulation without technical
          agents. The volume forms a steady flow with little
          deviations from the mean volume.}
        \label{fig:OrdersFlow_Fundamentals}
    \end{subfigure}

    \begin{subfigure}[b]{0.7\textwidth}
        \includegraphics[width=\textwidth]{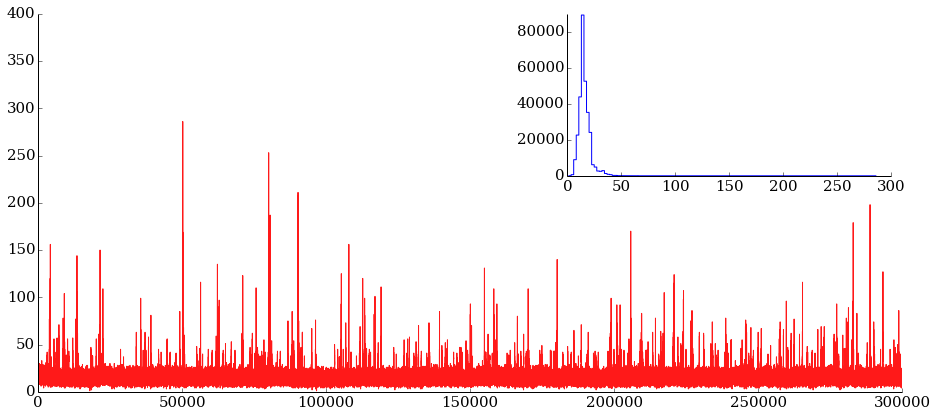}
        \caption{Trading volumes for a simulation with technical
          agents. There are big fluctuations in volume, rising above
          the "base line" created by the fundamental agents. These
          fluctuations are a consequence of the activity of technical
          agents.}
        \label{fig:OrdersFlow}
    \end{subfigure}
    \caption{Representative trading volumes for runs of the model
      without technical agents (a) and with technical agents (b). As
      can be seen, there are large fluctuations of the volume over
      time when technical agents are included in a simulation. The
      insets in each figure show the distribution of
      flows.}\label{fig:OrderFlows}
    
\end{figure}

This similarity in the central part of the volatility distributions in
the scenarios with and without technical agents, along with a similar
result obtained by Schmitt et al\cite{Thilo2012} with their model, in which the
agents place orders with exponentially distributed volumes, is of
interest since the flows of orders are very different in both cases
(see figures \ref{fig:OrdersFlow_Fundamentals} and
\ref{fig:OrdersFlow_Fundamentals}), yet, the majority of the
volatilities can be described by log-normal distributions. This
result suggests that the order book mitigates in some sense, the
variations in the shape of the incoming order "signal", in such a way
that the variations in price (the volatilities) are not strongly
affected by changes in the distribution of orders placed into the
book.

\begin{figure}[!htbp]

\centering
\includegraphics[width=0.7\textwidth]{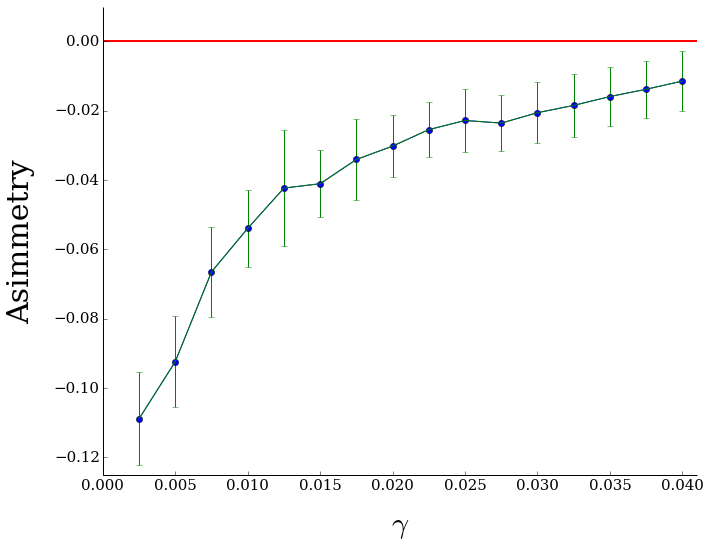}
\caption{Mean skewness of the distributions of returns as a function
  of the profit taking threshold $\gamma$. As $\gamma$ becomes
  smaller, more profit taking takes place and the mean skewness of the
  distributions of returns becomes more negative.}
\label{fig:Asymmetry_Ensemble}
\end{figure}

In Figure \ref{fig:Asymmetry_Ensemble} we plot the values of the
average skewness of an ensemble of 50 simulations (for every point in
the plot) as a function of the parameter $\gamma$. As explained above,
this parameter controls how often the population of technical agents
engage in profit taking. In the framework of our model, this behavior
is the cause of the asymmetry between losses and gains in the
distribution of returns, since by enanging in profit taking, the
population of technical agents creates large falls in the price of the
asset.

The mean skewness we measured in the empirical data obtained from
QuantQuote\cite{QuantQuote} has a value of $-0.33$;
close to the minimum average skewness obtained in our model with the
technical agents population engaging frequently in profit taking at
$\gamma=0.0025$. The number of companies with a skewness
within the interval $[-0.5,0]$ is $199$, which represents $39.8 \%$ of
the companies listed in the S\&P500. 

\begin{figure}[!htbp]
\centering
\includegraphics[width=0.7\textwidth]{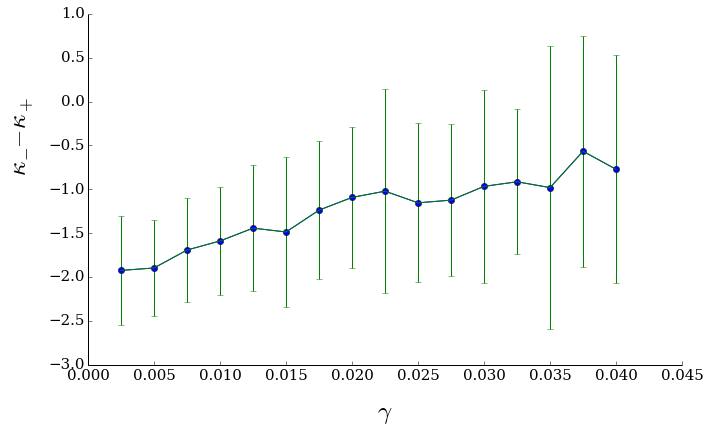}
\caption{Mean difference of the exponents $\kappa_-$ and $\kappa_+$ of
  the power law fits for the absolute value of the negative (red) tail
  and the positive (blue) tail of the distributions of returns.  This
  difference tends to decrease as $\gamma$ becomes larger, suggesting
  that the tails of the distribution tend to collapse one on top of
  the other as the technical agents engage less frequently in profit
  taking.}
\label{fig:Exponent_Differences}
\end{figure}

\begin{figure}[!htbp]
\centering
\includegraphics[width=0.7\textwidth]{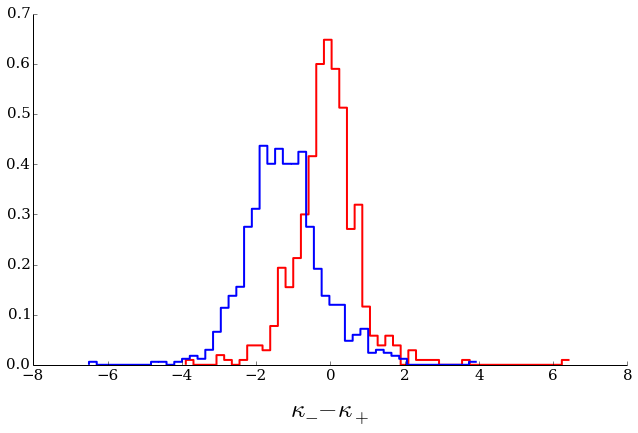}
\caption{Distribution of the differences between the negative and
  positive tail exponents of the powerlaw fit. The red histogram
  corresponds to the values of the empirical data, the blue one
  corresponds to the data generated by the model. A significant number
  of companies present values in the range generated by our
  simulations.}
\label{fig:Exponent_Differences_Distribution}
\end{figure}

In figure \ref{fig:Exponent_Differences} we present another test that
relates the asymmetry of the distribution of returns to the practice
of profit taking. In this figure we present the differences between
the exponents of the power law fits for both the positive tail and the
absolute value of the negative tail of the distribution of returns for
several ensembles of 50 simulations in which we varied the parameter
$\gamma$.  As can be seen in figure \ref{fig:Exponent_Differences},
we obtain mean values of the difference $\kappa_{-} - \kappa_{+}$ in a
range of $[-1.92,-0.56]$; the distribution of values for this
difference as measured in the empirical data is in the figure
\ref{fig:Exponent_Differences_Distribution}. The differences between the exponents for the power law fit obtained from the data generated with our model present a significant overlap with the empirical ones.

\begin{figure}[!htbp]
\centering
\includegraphics[width=0.7\textwidth]{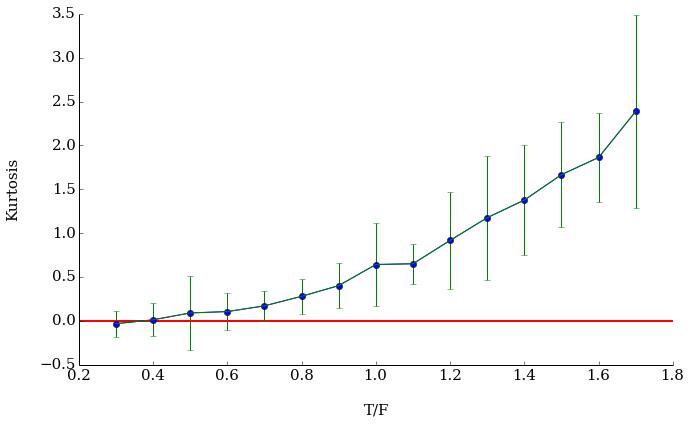}
\caption{Mean kurtosis of the distributions of returns as a function
  of the ratio of technical to fundamental agents. As the proportion of
  technical agents increases, so increases the mean kurtosis in the
  ensemble of simulations.}
\label{fig:Kurtosis_Ensemble}
\end{figure}

Similarly, in Figure \ref{fig:Kurtosis_Ensemble} we plot the average
kurtosis of an ensemble of 50 simulations as a function of the
fraction of technical agents in the population. The kurtosis shows an
increase with the number of technical agents, which strongly suggests
that they are responsible for the deviations from normal behavior
observed in the distribution of returns. Empirically, the kurtosis
measured on the various companies listed in the S\&P500 span a wide
range of values, with some companies having a kurtosis higher than
$100$. With our model, we were able to produce kurtosises as high as
$7$ when the population of technical agents was almost twice the size
of the population of fundamental agents. Unfortunately, using higher ratios 
without compromising the stability of the simulations requires a much larger 
total population of agents which is beyond our computational capacities.

\newpage

\section{Conclusions}

In this work we studied an agent based model of a single asset
financial market with agents employing simple heuristic rules, which
is capable of replicating the stylized facts reported in the
literature. As in the LM model\cite{Lux1999}, we divided the
population of agents into two groups according to the type of trading
strategy they use: fundamental agents and technical agents. Further,
we added heterogeneity within each group by varying the values of the
parameters that control each agent's behavior. Our aim was to create a
model whose agents behaved realistically, as in the LM model, but with
equally realistic market structures, namely, trading via a limit order
book.  We find, in accordance with previous models, that when the
population of agents include technical agents, the returns present
volatility clustering and a heavy tailed distribution. Further, we
found that essentially no autocorrelation of the returns was present
for any configuration of the populations. In addition to these main
stylized facts, we find that when we allow the population of technical
agents to engage in profit taking, the distribution of returns displays
negative skewness and an asymmetry between losses and gains
appears. By varying the frequency with which technical agents engage in
profit taking, we can generate return distributions with varying
degrees of separation in the tails. This dependence of the skewness
over the frequency of profit taking suggests that this practice may be
one of the causes of the appearance of the asymmetry in real financial
markets.

Regarding the distribution of volatilities we find that only its
central part is qualitatively similar to a lognormal distribution when
technical agents are included in the population. If, on the other
hand, we only include fundamental agents, the volatilities are
remarkably well described by a lognormal distribution. The similarity
of the volatility distributions in both scenarios, at least in the
central part, suggests that its shape may not be strongly dependent on
the detailed properties of the flow of incoming orders, since this
flow varies significantly when technical agents are inserted in the
population as compared with a population comprised entirely of
fundamental agents.

We accompany our results with empirical data from real financial
series chosen to illustrate the various stylized facts reproduced by our
model.

In its present state, the model represents a single asset market,
however, it is simple enough to be extended in several ways. For
instance, an interesting extension to the model would be to increase
the number of assets in the market and to limit the credit available
to each agent. By doing this, the well being of the different
``companies'' associated to the different assets could become
correlated depending on the shifts of the demand for each
asset. Thus, we could inquire into the nature of these correlations,
and how they are related to the composition of the population of
agents. Another interesting modification would be the introduction of
sequences of catastrophic news. The model will allow us to
study how fast and in which way the market recovers to states observed
previous to the arrival of the catastrophic news, if it recovers at
all, and if the composition of the population affects this recovery.

\ack
R.M. acknowledges financial support from CONACyT. We thank
Stephanie Rend\'on, Ana Contreras and Paulino Monroy for useful discussions.

\clearpage
\section*{References}
\bibliographystyle{unsrt}
\bibliography{References}
\end{document}